# Rate-Dependent Reversibility and Lithium Losses in Hybrid Anode–Collector Metal Electrodes

Arturo Galindo[a], Jesús Díaz-Sánchez[b,c], Sunil Kumar[d], Bouthayna Alrifai[d], Andrea Marchetti[a], Gastón García[d], Celia Polop[b,c,e] and Enrique Vasco[a*]

a.  Instituto de Ciencia de Materiales de Madrid, Consejo Superior de Investigaciones (CSIC), Spain
b.  Departamento de Física de la Materia Condensada, Universidad Autónoma de Madrid, Spain
c.  Instituto Universitario de Ciencia de Materiales Nicolás Cabrera (INC), Universidad Autónoma de Madrid, Spain
d.  Centro de Micro-Análisis de Materiales (CMAM), Universidad Autónoma de Madrid, Spain
e.  Condensed Matter Physics Center (IFIMAC), Universidad Autónoma de Madrid, Spain

Dated: March 12, 2026

**Corresponding author:**

Enrique Vasco

Instituto de Ciencia de Materiales de Madrid

Consejo Superior de Investigaciones Científicas

CL. Sor Juana Inés de la Cruz 3, Cantoblanco

28049 Madrid, Spain

phone: +34-913348981 / email: enrique.vasco@csic.es

**Abstract**

Understanding how practical lithium storage capacity varies with charge–discharge rate is crucial for designing durable anode-free lithium batteries. We examine the lithiation behavior of single-element metal electrodes—Al (alloying), Mg (solid-solution intercalation), Ag (solid-solution then alloying), and Cu (surface Li plating)—to determine how their mechanisms influence reversibility, measured by coulombic efficiency. Using electrochemistry combined with depth-resolved ion-beam profiling, we map local coulombic efficiency across current densities and identify dominant lithium-loss pathways. Ag uniquely sustains fast kinetics and high reversibility at elevated rates due to rapid formation of γ-brass–type alloy phases. In contrast, Mg and Al show increasing irreversibility from kinetically or structurally driven Li trapping, while Cu exhibits the largest losses through porous, highly reactive plated lithium. These results reveal fundamental limits of anode-free systems that depend on reversible Li plating without excess lithium and underscore the importance of metal selection for stable, high-rate performance.

**TOC Graphic**

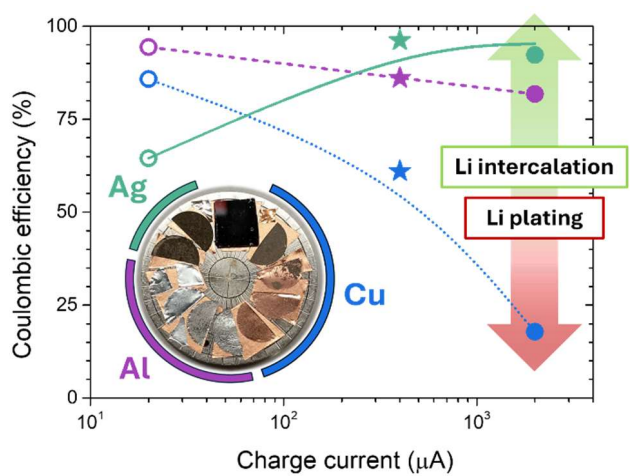

**TOC Caption:** Effect of charge-discharge rate on reversibility for the different investigated lithiation mechanisms: Li-metal alloying (represented by Al), solid-solution intercalation followed by alloying (Ag) and Li plating (Cu). Ag shows the best high-rate performance via rapid Li intercalation into gamma brass–type Ag-Li alloys, whereas Al (and Mg) is limited by structural changes and Cu suffers extensive Li plating and irreversible Li losses.

The use of alkaline-ion batteries in energy-intensive applications, such as electric mobility, requires optimizing the balance between specific capacity (or capacity density, depending on whether the constraints are weight or volume) and charge/discharge rates.[1] This balance defines the duty factor, power delivery, and practicable capacity of the battery. High charge rates can lead to capacity fading by triggering irreversible processes such as secondary reactions, phase changes, metal plating, increased internal resistance, and heat dissipation. On the other hand, high discharge rates prevent the battery from accessing energy stored in slow-kinetic electrochemical processes. For instance, although $LiCoO_2$ is known for its high specific lithium storage capacity, its slow Li-ion kinetics restrict its specific power at low C-rates, making it unsuitable for demanding applications like electric vehicles. In general, increasing the charge and discharge rates leads to a reduction in the practicable capacity of the battery.[2,3]

In this context, battery manufacturers define nominal capacity as the amount of charge (or energy, when multiplied by the battery's operating voltage) that a fully charged battery delivers under model (mostly impractical) operating conditions.[4] For instance, Li-ion batteries (LIBs) are typically charged using a two-step protocol: an initial gross charging at constant current (CC) $I_C$ until reaching the maximum operating voltage, followed by a time-consuming charge saturation regime at constant voltage (CV).[5] As higher $I_C$ is used, the first step shortens; however, more time is required for saturation capacity, resulting in a net increase in charging time by the CC-CV procedure. Since charging time is a key parameter for electric vehicle (EV) consumers, several commercial solutions have been adopted so far: (i) Oversizing the battery to store sufficient energy using exclusively the CC regime, while underutilized portion serves as a safety margin to replace degraded cells.[6] This strategy is supported by manufacturers, while customers remain unaware of the energy management approach. (ii) Consumers are encouraged to avoid full charges and deep discharges, argued to reduce degradation outside the 20–80% SOC window. As a result, the consumers pay for the battery's full capacity but cannot fully use it under the manufacturer's recommended "responsible-use" guidelines.[7] (iii) Using increasingly complex modified CC-CV protocols (e.g., gradually reducing $I_C$ at high SOC to limit degradation), which raises the cost of EV chargers and makes them brand-specific.[8] As long as EV batteries require significantly longer charging times than refueling combustion vehicles, fully electric mobility will struggle to serve as a globally reliable alternative.[9] Consequently, hybrid architectures like extended-range electric vehicles that integrate a small fuel engine to recharge the battery are gaining substantial market share.[10]

In this work, we investigate how charge–discharge rate affects reversibility in Li-ion batteries using metal electrodes that integrate both anode and current-collector functions (Hybrid Anode-Collectors, HACs). Implementing HACs can substantially increase cell-level energy density and simplify device architecture by reducing the number of stacked components. We examine four single-element metals (Al, Mg, Ag, and Cu) selected for their distinct lithiation mechanisms: Al rapidly forms Li–Al alloys; Mg accommodates Li through a solid-solution mechanism accompanied by structural changes that slow down kinetics; the Ag lattice hosts up to ~35 at.% Li before transitioning to intermetallic phases; and Cu remains essentially immiscible with Li, acting primarily as a barrier and promoting surface plating.[11]

**Figure 1** shows the GCD voltage profiles of the first electrochemical cycles for each metal component, with curves transitioning from black to red as the cycling progresses, at different discharge/charge (lithiation/delithiation) currents (namely, $I_D = I_C =$20 μA and 2 mA, left and right columns) under a C/2 rate. Each row corresponds to a different metal electrode. Additional intermediate-current data can be found in Section S1 of Supporting Information (SI). Since pure metals exhibit higher open-circuit voltages ($V_{oc}$ vs. Li/Li$^+$) than the lithiated counterparts, each electrochemical cycle begins with the discharge-lithiation of metals (solid curves in Fig. 1) followed by the charge-delithiation (dashed curves). The GCD curves show distinct behaviors linked to different lithiation mechanisms: (i) voltage plateaus above zero signal intermetallic alloy formation (phases indicated by green labels); (ii) pseudolinear voltage-capacity relationships indicate solid solution development with Li intercalation (phases indicated by pink labels); and (iii) plateaus at $V_l = V_d = 0$ represent Li plating. Phase identification (green/pink) integrates GCD data (mainly delithiation curves) and cyclic voltammetry (not shown here), interpreted with binary phase diagrams for Li-metal systems.[12–15] The intersection of each delithiation curve with the upper axis at 3.0 V estimates coulombic efficiency ($\varepsilon_c$) for each cycle, with the absence of an intersection corresponding to 100% (e.g., 20 μA-cycled Mg). The behavior identified for each element is described in detail below.

The lithiation curves for **Al component** (first row in Fig. 1) show a drop from $V_{oc} \approx 2.4\ V$ to a lithiation plateau $V_l \approx 0.3\ V$, with a nucleation barrier of 0.1 V (feature A), followed by a gradual decline toward 0 V at higher capacities (feature B). The plateau is due to Li reduction in the $LiAl$ alloy, and the nucleation barrier reflects initial wetting-like Li accumulation on Al sufficient for absorption but not plating. The decreasing trend in feature B likely results from forming Li-richer Li–Al alloys[13] (such as $Li_3Al_2$, $Li_{2-x}Al$ and $Li_9Al_4$) and potential Li plating. The shift in the capacity at which $V_l$ drops to zero with lithiation current (e.g., >3 mAh at $I_C = 400$ µA in Fig. S1 (SI) and >1 mAh at 2 mA) indicates that lithiation kinetics govern this behavior. Thus, if the lithiation current exceeds the rate of alloying front propagation within Al, rapid transitions between Li-rich alloys occur until excess Li plates on the Al surface. Alloying front progression, rather than Li diffusion (which is faster in $Li-Al$ alloys than in pure Al), chiefly limits lithiation rate.[16] Local formation of Li-rich alloys is best seen in delithiation curves at $I_D = 2$ mA (features B'), as explained below.

In the case of **Mg component** (second row in Fig. 1), as Mg and Li are fully miscible, an intercalation solid solution forms across all compositions,[14] giving Mg a high theoretical Li storage capacity. At low current (20 µA), the lithiation/delithiation GCD profiles reveal a gradual transformation of the host *hcp* Mg lattice (space group $p6_3/mmc$) into a disordered *bcc* $Im\bar{3}m$ lattice as Li increases from 18 to 35 at.% (feature C'). This transition reflects a change from Li intercalated in *hcp* Mg ($\alpha Li_x Mg_{1-x}$) to Mg incorporated into *bcc* Li ($\beta Li_x Mg_{1-x}$). At higher current (2 mA), the Mg delithiation curves (dashed lines) become fuzzy (feature C') when Li drops below 35 at.%. This is due to the kinetics of delithiation at 2 mA and phase transformation occurring on similar timescales, causing sudden voltage changes as Mg intermittently releases more Li.

For **Ag component** (third row in Fig. 1), at low current (20 µA) the GCD curves reveal the formation of an intercalation solid solution $\alpha Li_x Ag_{1-x}$ that holds up to 35 at.% Li at room temperature within the fcc $Fm\bar{3}m$ Ag lattice. As Li increases (35–50 at.%), a cubic $Pm\bar{3}m$ $LiAg$ alloy (with $V_d = 0.33$ V) is formed. For Li above 65.5 at.% (2 mA), the delithiation voltage profile is pseudolinear, resembling solid-solutions behavior, although the binary phase diagram does not predict such solutions at these compositions. Instead, this may reflect rapid transitions between related intermetallic alloys, blending expected plateaus into a pseudolinear trend. Literature[15] reports that, for 65.5–92 at.% Li, gamma brass-type $\gamma Li_y Ag_z$ alloys (specifically $\gamma_3\ Li_9 Ag_4$, $\gamma_2\ Li_4 Ag$ and $\gamma_1 Li_9 Ag$, all with cubic $I\bar{4}3m$ structures) are formed. Their GCD profiles (labeled green) are discussed in detail in Fig. 2.

And in the case of **Cu component** (fourth row in Fig. 1), the *fcc* $Fm\bar{3}m$ Cu lattice admits up to 21 at. % Li as an intercalation solid solution $\alpha Li_x Cu_{1-x}$,[15] but excess Li forms metallic deposits (with $V_l = 0\ V$) on the Cu surface due to limited miscibility at higher concentrations. During delithiation, Cu shows mild fuzzing in 2-mA delithiation curves (feature D'), likely from Li plating restructuring. This effect is weaker than in Mg delithiation at high currents and disappears at lower charging currents (see Section S1).

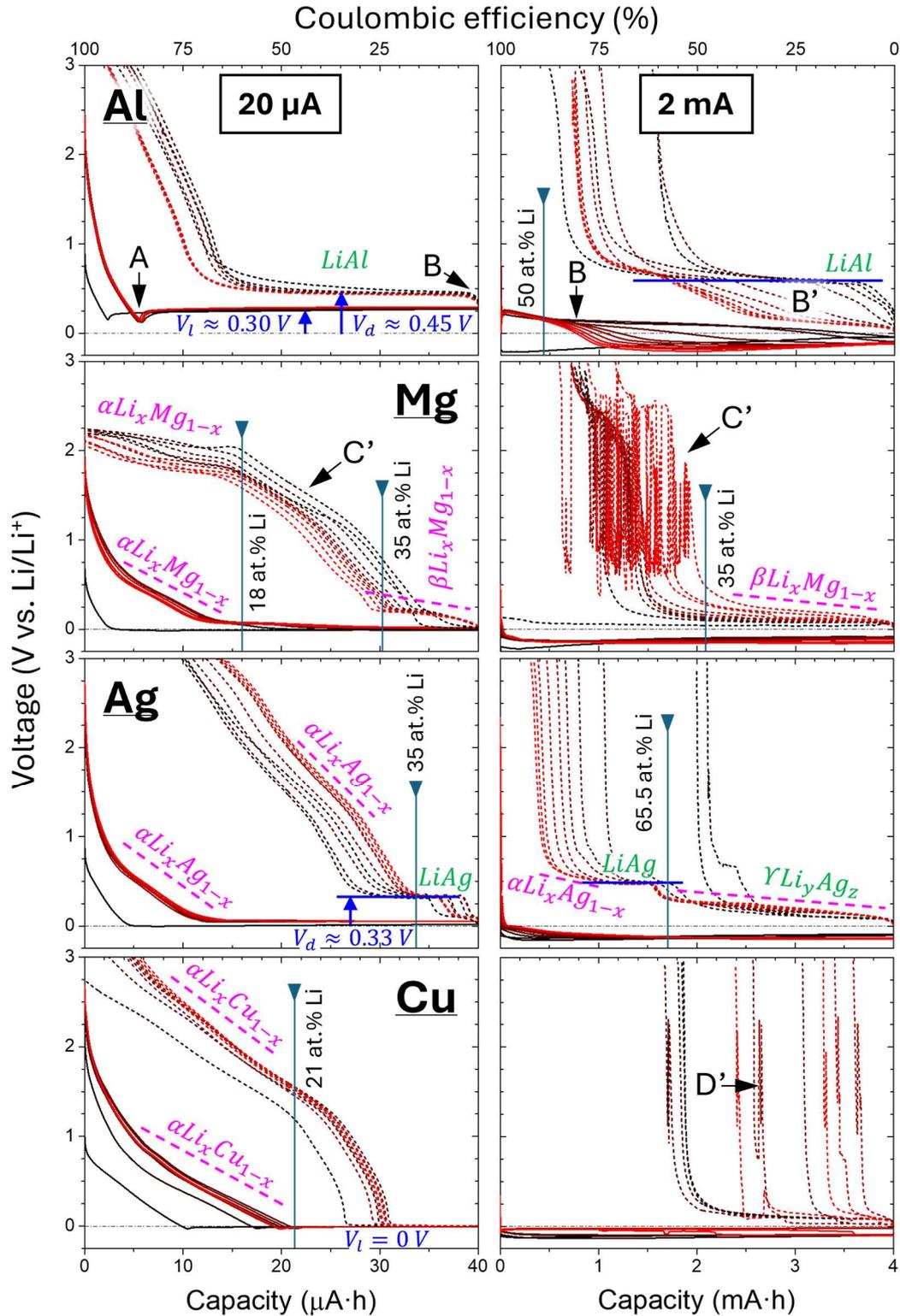

**Fig. 1** GCD voltage profiles for the first electrochemical cycles of the metal components, measured at different delithiation/lithiation currents (with $I_C = I_D$) under a C/2 rate. Green, pink and blue labels denote the formation of intermetallic alloys, intercalation solid solutions, and their respective electrochemical potentials. Features A, B, C, and D (with/without apostrophe to distinguish lithiation and delithiation events) are detailed in the main text. The intersection of each charging curve (dashed lines) with the upper voltage axis estimates coulombic efficiency for each cycle, with the absence of an intersection corresponding to 100% (e.g., 20 μA-cycled Mg).

To accurately interpret the different processes occurring during the cycling of Al and Ag electrodes, GCD experiments were performed using dissimilar charging and discharging currents. Lithiation at 2 mA produced high Li content, allowing the examination of Li-rich alloys, while delithiation at 400 μm enabled clearer observation of alloy phase transitions. **Figure 2** displays voltage profiles with distinct plateaus and transients representing specific alloys and their transitions.

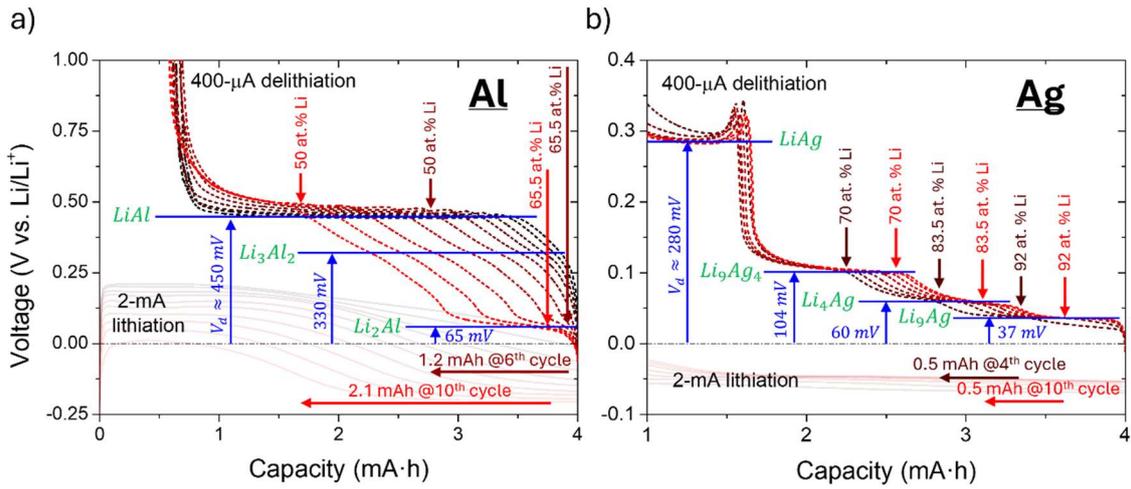

**Fig. 2** GCD voltage profiles for lithiation (weak, solid curves) and delithiation (intense, dashed curves) of (a) Al and (b) Ag at different discharging ($I_D$ =2 mA) and charging ($I_C$ =400 μA) currents. The color convention is like that in Fig. 1.

Figure 2a shows that, for Al, the delithiation voltage profiles broaden with cycling, and the capacity needed to transition from the $Li_2Al$ alloy (65.5 at.% Li) to the $LiAl$ alloy (50 at.% Li) increases from 1.2 mAh in the sixth cycle to 2.1 mAh in the tenth. This 75% increase in required capacity indicates a substantial lithiated volume growth between cycles, confirming that lithiation front propagation is the rate-limiting step in Al. In contrast, for Ag in Fig. 2b, required capacities for transitions between gamma brass-type alloys remain constant regardless of cycles (e.g., the transition from $Li_9Ag$ to $Li_4Ag$ always requires extracting 0.5 mAh of stored Li in our systems). This suggests a stable lithiated volume and pointing to Li diffusion as the main rate-limiting factor in Ag.

**Figure 3** shows the evolution of coulombic efficiency ($\varepsilon_c$) as a function of cycle number for different discharge/charge currents. The values of $\varepsilon_c$, determined from the intersections of the delithiation curves with the upper axis in Figs. 1 and 2, capture both continuous and intermittent Li release during metal delithiation (the latter being particularly relevant for Mg, see SI). Generally, as outlined in the introduction, higher cycling currents result in lower $\varepsilon_c$, especially when charge/discharge rates outpace electrode kinetics. Curves with open symbols (corresponding to lower $I_C$) in Fig. 3a exhibit higher efficiencies compared with closed symbol curves (higher $I_C$), indicating a lower contribution of irreversible processes at reduced current. Lithium storage by pure plating, as in Cu (blue solid symbols), shows the greatest drop in efficiency, whereas coulombic efficiencies in alloy systems (e.g., Al) or intercalation solid solutions (Mg) are less affected. Ag behaves uniquely: after initial cycles, its efficiency rises with increasing current once $LiAg$ alloy capacity is surpassed and Li stores mainly in gamma brass-type $\gamma Li_y Ag_z$ alloys. Additional increases in current then slightly reduce efficiency (Fig. 3b).

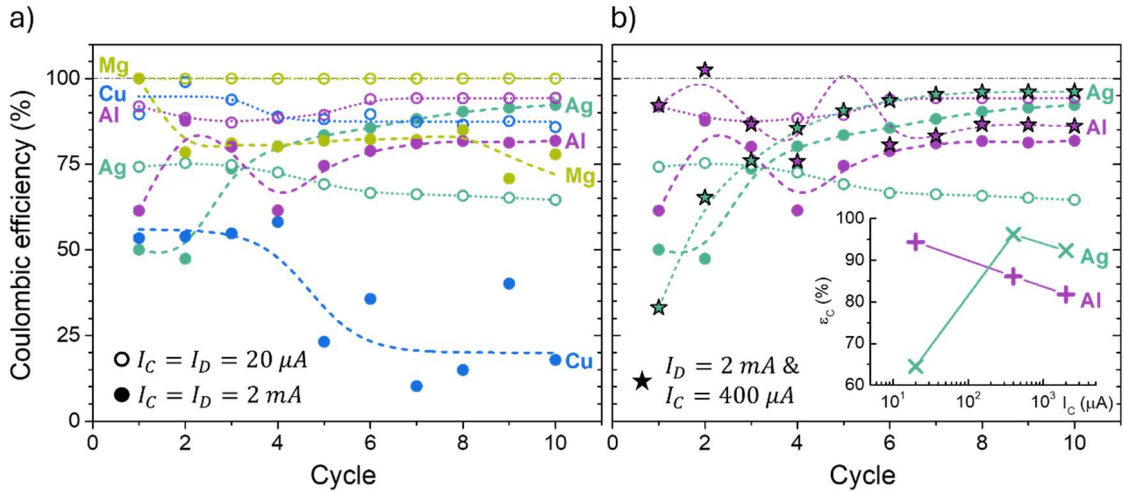

**Fig. 3** Cyclic coulombic efficiencies ($\varepsilon_c$) for the investigated metals at different charge/discharge currents: (a) $I_C = I_D$ and (b) $I_C = I_D/5$. Al and Ag curves are repeated on (b) for comparison. $\varepsilon_c$ is derived from the intersection of the delithiation curves (dashed ones) with the upper axis (at 3.0 V) in Figs. 1 and 2. [inset in (b)] $\varepsilon_c$ as a function of charging current after ten cycles.

Li depth profiles of metal components were analyzed by ion beam analysis (IBA) after cycling to selected states of charge to identify regions where Li becomes irreversibly trapped under various charging and discharging conditions. Profiles were generated by fitting 3-MeV H$^+$ NRA and 4-MeV He$^+$ RBS spectra with MultiSIMNRA 1.6,[17] and profile areas were normalized to electrochemically exchanged Li. This approach provides Li sensitivity ~2.5·10$^{16}$ at./cm$^2$ [<0.45 at.% for our sample geometry and composition] comparable to NRA (using the collision cross sections reported by Paneta et al.)[18] and a ~60 nm depth resolution similar to He$^+$ RBS (as predicted by the RESOLNRA 1.11 code[19] for our sample geometry and composition). Related spectra are shown in Section S2 (Figs. S2 and S3).

**Figure 4** shows Li depth profiles, $[Li] = f(depth)$, resulting from discharge-lithiation ($[Li]_l$, blue curve) and charge-delithiation ($[Li]_d$, green curve) of metal components after ten electrochemical cycles at two different currents ($I_C = I_D = 20$ μA and 2 mA). The magenta curve (right axis) tracks cumulative coulombic efficiency by comparing extracted Li ($[Li]_l - [Li]_d$, blue-filled area) to intercalated/plated Li ($[Li]_l$, blue curve). Each graph also reports integral Li losses derived from the ratio of the areas under the $[Li]_d$ and $[Li]_l$ profiles. When interpreting tenth-cycle Li profiles, note that Li exchange with the metal bulk occurs via the surface, producing local accumulations where Li transport diverges.

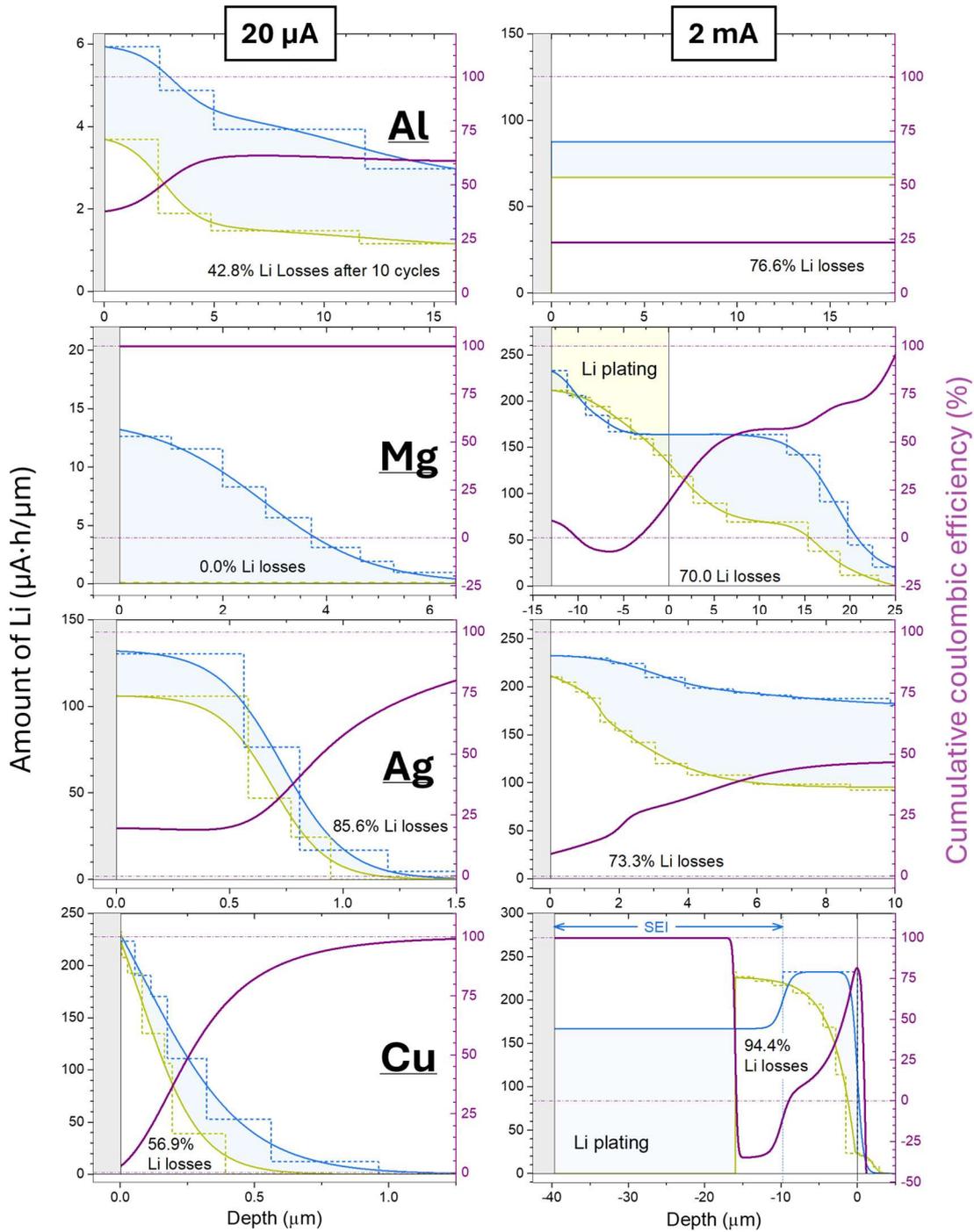

**Fig. 4.** Li depth profiles after ten cycles, measured during discharge-lithiation (blue curve) and charge-delithiation (green curve) at two different charge/discharge currents ($I_C = I_D$ =20 μA and 2 mA). Profiles were obtained by simulating the IBA spectra (see SI) with the MultiSIMNRA 1.6 code and normalized to the total electrochemically exchanged Li. The magenta curves (right axis) display the cumulative coulombic efficiency as a function of depth. Li losses for each metal and current are indicated in the respective graphs. Definitions of these parameters are detailed in the main text.

The close similarity between the shape of the lithiation and delithiation Li depth profiles in **Al component** (first row in Fig. 4) for both regimes indicates that irreversible Li is distributed uniformly throughout the Al volume (consistent with the nearly constant cumulative coulombic efficiency for depths >4 µm). This suggests that the irreversible component arises from structural changes created as the lithiation front advances under compressive stress caused by alloy volumetric expansion. Li tends to migrate out of these compressed regions, which can slow down the progression of the lithiation front and make it the rate-limiting step.[20] When the front encounters defects that partially relax the volumetric strain, lithiation is locally accelerated; however, such defects can also trap Li in irreversible or slow-diffusing states. This trapped Li remains in nanoporous Al after delithiation, leading to faster lithiation in subsequent cycles, as shown in Figure 2. Conversely, at low current cycling (20 µA), some adsorbed Li remains exposed to the electrolyte for longer periods and undergoes irreversible side reactions, which explains the reduction in cumulative efficiency near the Al surface.

The **Mg component** (second row in Fig. 4), at low charge/discharge currents (20 µA), exhibits no detectable irreversible Li accumulation (the green profile is essentially zero). When the current is increased to 2 mA, however, Li begins to accumulate at the electrode surface because Mg lithiation becomes sluggish once the Li content reaches approximately 18–35 at.%. In this composition range, the Mg lattice undergoes structural transitions between *hcp* and *bcc* phases, which slows Li insertion kinetics. As a result, excess Li plates onto the surface, where it reacts irreversibly with the electrolyte to form a solid–electrolyte interphase (SEI). This process produces a near-zero cumulative coulombic efficiency at the Mg surface (Fig. 4) and causes the net efficiency after ten cycles to drop from 100% at 20 µA to 72% at 2 mA (Fig. 3a).

In the case of **Ag component** (third row in Fig. 4), at low charge/discharge currents (20 µA), Li penetrates only shallowly (note the much smaller depth range of 1 µm compared with 16 µm in Al and 6 µm in Mg). The relatively similar lithiation and delithiation depth profiles indicate that a fraction of subsurface Li is uniformly trapped in irreversible states. When the current is increased to 2 mA, (i) Li penetrates significantly deeper, surpassing the depth reached in Mg at the same current (compare the Li content at 10 µm, the limit of IBA detectability in Ag: 180 µAh/µm for Ag and 162 for Mg); and (ii) the lithiation and delithiation profiles differ clearly near the Ag surface. Both observations (i and ii) are explained by the divergence in Li transport as the Li–Ag system crosses alloy-phase regions with distinctly different diffusivities (details in Section S3). Recent studies[21] indicate that the moderate Li diffusivity in pure Ag (~$10^{-7}$ cm$^2$/s, typical of intercalation-type solid solution $\alpha Li_xAg_{1-x}$) decreases down to ~$10^{-9}$ cm$^2$/s as the intermetallic $LiAg$ phase begins to form above ~35 at.% Li, before increasing again to the range of $10^{-6}$–$10^{-5}$ cm$^2$/s at higher Li loadings (65.5–92 at.% Li), where Li is incorporated into the gamma brass-type $\gamma Li_yAg_z$ alloys. As a consequence, lithiation/delithiation kinetics slow markedly within regions dominated by the $LiAg$ alloy, specifically when the local Li concentration lies between ~35–65.5 at.% (the case for the 20 µA depth profiles). The divergence in Li transport accounts for the above observations for 2 mA: (i) At higher currents, Li rapidly reaches concentrations above ~65.5 at.%, enabling storage within the high-diffusivity $\gamma Li_yAg_z$ alloys. This accelerates Li transport and leads to significantly deeper lithiation. (ii) During delithiation, once the local Li content drops below the ~65.5 at.% threshold, transport slows as the material re-enters the low-diffusivity $LiAg$ alloy regime. This leads to an excess of Li, relative to the lithiation profile, becoming trapped in slow-kinetic states near the Ag surface. During the subsequent lithiation step, part of this accumulated Li is gradually redistributed into the Ag volume. It is important to note that the Li depth profiles in Fig. 4 correspond to IBA measurements collected after ten cycles, during which irreversible or slow-kinetic Li progressively accumulates (as shown in **Figure 5**). Consequently, any substantial differences between the lithiation and delithiation profiles (beyond point-level noise) should be interpreted as signatures of repeated back-and-forth transport processes (discussed in Section S3 of SI).

For **Cu component** (fourth row in Fig. 4), limited miscibility between Cu and Li causes Li to accumulate on the Cu surface, producing poor reversibility similar to Li plating directly exposed to the electrolyte. Reversibility decreases sharply with increasing current, falling from 86% at 20 µA to 20% at 2 mA (Fig. 3a), as faster deposition promotes greater surface roughening and porosity. This morphology, governed by a diffusion-limited aggregation (DLA) mechanism, increases the exposure of Li to irreversible reactions (including SEI formation, Li oxidation, and the development of high-surface-energy coatings) regardless of whether the electrolyte is liquid or solid-state. The low Li self-diffusivity in Cu (~$10^{-11}$ cm$^2$ s$^{-1}$)[22] is insufficient to counteract DLA-induced roughness, especially at high currents, exacerbating Li loss. Only during delithiation, the Li plating undergoes restructuring and partial densification, as reflected in Fig. 4 by the region of negative cumulative efficiency. The stochastic nature of DLA-driven electrodeposition explains the large dispersion in coulombic efficiencies observed for Cu at 2 mA (Fig. 3a).

To correlate the electrochemical response with the IBA results, we monitor the evolution of Li content (capacity, Q) over ten cycles at a 0.5C rate. **Figure 5** shows the amount of Li stored in the metal components during discharge–lithiation (blue bars) and the residual Li remaining after charge–delithiation (green bars, $Q_r$). Stored Li includes short-term reversible, long-term reversible, and irreversible fractions. Short-term reversible Li refers to the active material that is intercalated and deintercalated within a single cycle. In contrast, long-term reversibility corresponds to Li that is gradually released over multiple cycles due to slow delithiation kinetics. This manifests as delithiation capacities that exceed the lithiation capacity of the preceding cycle, a behavior commonly observed during early cycling. To describe residual Li ($Q_r$), we apply a power-law model dependent on cycle number $n$: $Q_r \sim (n - n_0)^\beta$. The exponent $\beta$ characterizes irreversibility: $\beta = 1$ means uncorrelated cycles, where Li losses accumulate linearly ($Q_r \sim [100\% - \langle \varepsilon_c \rangle]n$, with $\langle \varepsilon_c \rangle$ being the average coulombic efficiency); $\beta < 1$ a self-limiting process with $Q_r$ tending to saturate over time; and $\beta > 1$ a sustained correlation between cycles, with irreversibly trapped Li promoting further losses in subsequent cycles (details in Section S4 of SI). The onset parameter $n_0$ indicates cycle-to-cycle correlation, with $n_0 \to 0$ for fast delithitation. Figure 5 presents power-law fits (dashed red lines) and Table I lists the fit parameters.

Increasing the charge/discharge currents generally boosts the fraction of lithium engaged in the electrochemical processes described by the parameters $n_0$ and $\beta$. Under higher C-rates, slow but reversible Li-exchange pathways may be shifted into later cycles (observed as an increase in $n_0$) or may instead become effectively irreversible, reflected by a rise in $\beta$. Both effects typically contribute to a measurable loss of battery capacity as the operating current increases.

For the **Al component** (first and second row in Table I), the fraction of irreversibly absorbed Li increases because slow-kinetic reversible Li transitions to an irreversible state as the current rises (i.e., $n_0$ decreases from 0.82 to approximately 0, while $\beta$ increases from 0.63 to 0.91). As shown In Fig. 4 (first row), IBA reveals that this irreversible Li is uniformly distributed throughout the Al volume, likely becoming anchored at structural defects that promote lithiation. The relatively low proportion of such defects compared with the density of alloy-forming sites explains why Li losses in Al remain self-limiting ($\beta < 1$) even though the degree of irreversibility increases. This self-limiting behavior accounts for the relatively high coulombic efficiency ($\varepsilon_C > 80\ \%$, Fig. 3b) maintained during Al lithiation at high currents.

In the case of **Mg component** (third and fourth row in Table I), no Li losses ($\beta = 0$) are detected during 20 $\mu A$-cycling. However, when the current increases to 2 mA, (the contribution of the slow-kinetic reversible fraction) $n_0$ rises to 0.84 while (that of the irreversible fraction) $\beta$ increases to 1.05, as determined from the upper intercepts of the delithiation curves (including the fuzzy regions) in Fig. 1 (second row). If, alternatively, coulombic efficiencies are estimated considering only the pre-fuzzy delithiation flanks (see Section S5 of SI), thereby effectively excluding slow delithiation, $n_0$ approaches zero and its contribution is reassigned to irreversible Li, resulting in a higher $\beta$ value of 1.45. In Fig. 4, IBA reveals that most of the irreversible Li corresponds to plated Li associated with SEI formation, which explains why Li losses become self-sustaining at higher currents ($\beta > 1$). Increased Li plating amplifies the specific surface area exposed to the electrolyte. Nevertheless, not all Li ends up as plating: a significant fraction remains adsorbed on Mg as an intercalation solid solution (Fig. 4). This coexistence of plating and intercalated Li accounts for the still-moderate coulombic efficiency observed during high-current cycling ($\varepsilon_C = 72\ \%$ at 2 mA, Fig. 3a).

Unlike Mg (which mainly forms solid solutions with Li) and Al (which forms Li alloys), **Ag component** (fifth and sixth row in Table I) is capable of both forming solutions and alloying, and demonstrates a unique behavior: its reversibility improves as the charge/discharge current increases. Specifically, the power-law exponent $\beta$ decreases from 1.16 at 20 μA to 0.43 at 2 mA, reflecting a shift from sustained Li losses (primarily arising from subsurface absorption into the slow-kinetics $LiAg$ alloy) to self-limiting Li losses associated with fast-kinetics brass-type $YLi_yAg_z$ alloy formation. Despite this enhanced reversibility at higher currents, a fraction of the recovered Li still exhibits slow kinetics, as indicated by the increase in $n_0$ from 0 to 0.78. This fraction is expected to influence the back-and-forth behavior detected in the Li depth profiles by IBA, as well as the system's overall efficiency over successive cycles, after its electrochemical response has stabilized for $n \gg n_0$.

**Cu component** (seventh and eighth row in Table I) exhibits sustained irreversibility that becomes more pronounced as the charge/discharge current increases. This trend is reflected in the rise of the power-law exponent $\beta$ from 1.06 at 20 μA to 1.15 at 2 mA. Notably, there is no meaningful contribution from slow-kinetics reversible Li to this behavior. The sharp decline in coulombic efficiency (from 86 % to 20 % in Fig. 3a) indicates that, once the limited surface miscibility of Cu with Li (~21 at.% Li) is exceeded, Li deposits exclusively as rough, porous metallic plating.

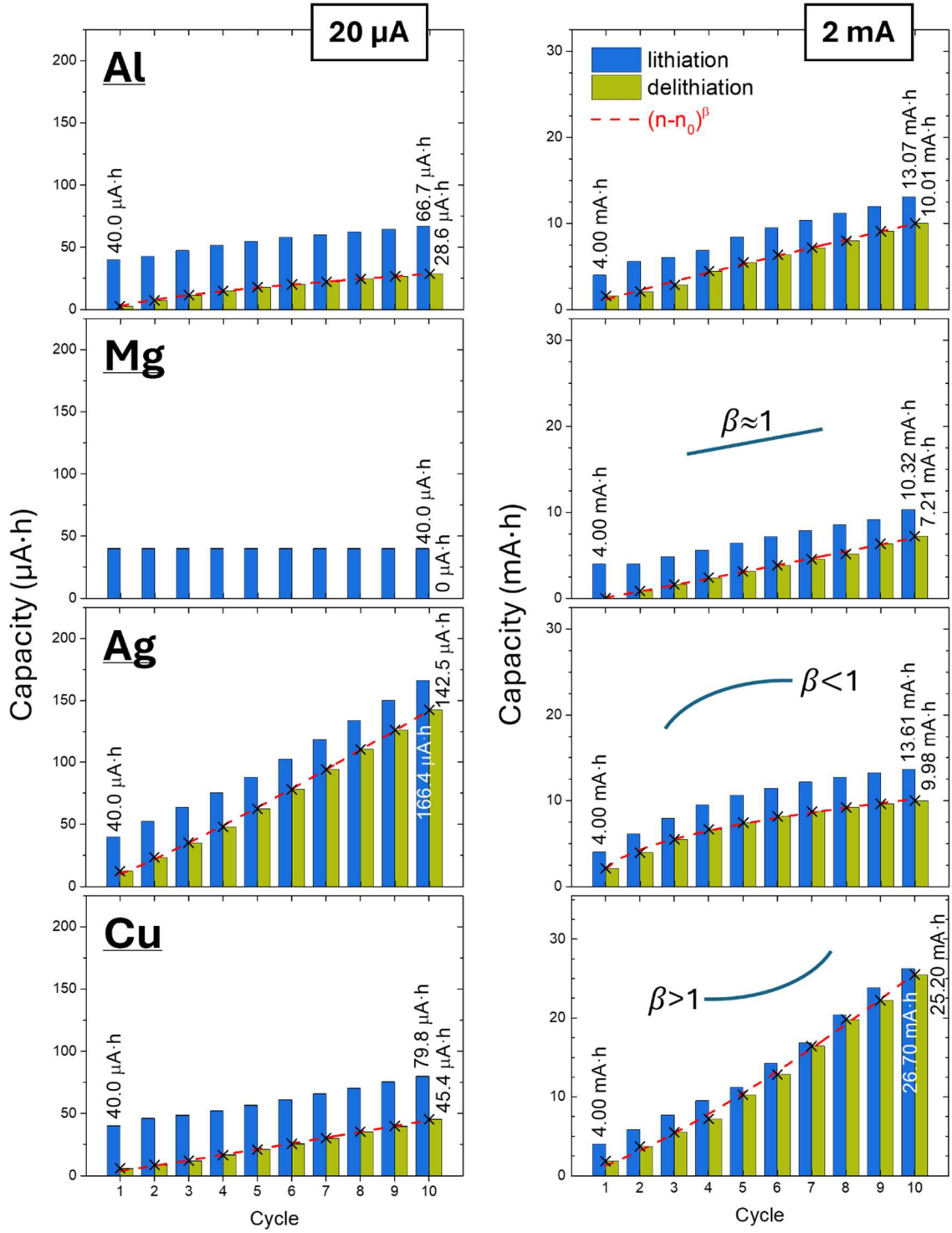

**Fig. 5.** Evolution of Li capacity in metal components over ten cycles, showing the amounts stored during discharge-lithiation (blue bars) and residual Li after charge-delithiation (green bars) at different charge/discharge currents. Dashed red curves illustrate power-law fits modeling the dependence of residual Li on cycle number. Each graph indicates the total Li capacity retained by the component after 10 cycles, with fitting parameters summarized in Table I.

| Metal – $I_C=I_D$ | $n_0$ | $\beta$ |
|---|---|---|
| Al – 20 µA | 0.82 ± 0.04 | 0.63 ± 0.02 |
| Al – 2 mA | 0 | 0.91 ± 0.03 |
| Mg – 20 µA | - | 0 |
| Mg – 2 mA pre-fuzzy* | 0.84 ± 0.23 0 | 1.05 ± 0.06 1.45 |
| Ag – 20 µA | 0 | 1.16 ± 0.02 |
| Ag – 2 mA | 0.78 ± 0.07 | 0.43 ± 0.02 |
| Cu – 20 µA | 0 | 1.06 ± 0.09 |
| Cu - 2 mA | 0 | 1.15 ± 0.03 |

**Table I.** Fitting parameters for power-law dependence $Q_r \sim (n - n_0)^\beta$, describing the evolution of residual capacity (Li losses) in each metal component during cycling. The physical interpretation of $\beta$ and $n_0$ is discussed in the main text.

*Note: The evolution of Li content in Mg under fast delithiation (pre-fuzzy regime) is detailed in Section S5 of SI.

Our analysis identifies two distinct mechanisms of Li loss in metal electrodes, each captured by the power-law exponent $\beta$, which quantifies how the amount of trapped Li evolves during cycling. When $\beta < 1$, Li losses are self-limiting: only a minority of sites (typically structural defects within the metal bulk that locally promote lithiation) irreversibly anchor Li or trap it in slow-kinetic states. In contrast, $\beta > 1$ signifies self-sustaining Li losses, with the number of trapping sites increasing from cycle to cycle. This regime is commonly linked to the growth of Li plating surface area, particularly when deposition follows a diffusion-limited aggregation (DLA) mechanism, which produces highly ramified, reactive Li structures. Such plating, whether interacting with liquid or solid electrolytes, continuously expands the active surface available for side reactions, thereby accelerating Li loss.

Among the metals studied, Ag demonstrates superior kinetics and reversibility for Li storage at high charge/discharge rates. This is attributed to Li intercalation into the Ag lattice, forming rapidly transitioning gamma brass-type alloys that exhibit electrochemical voltage profiles similar to those of intercalation solid solutions, but with much greater capacity. In contrast, Mg and Al undergo significant structural rearrangements during lithiation, which slows their reaction kinetics and leads to reduced coulombic efficiencies at high currents. These transformations introduce slow-kinetic pathways and irreversible trapping mechanisms that become increasingly pronounced under fast-cycling conditions. Finally, Cu shows the greatest Li losses, as Li is stored almost exclusively through surface adsorption and deposition as rough, porous metallic plating. This morphology strongly promotes side reactions and continuously increases the reactive surface area, underscoring the importance of minimizing or avoiding Li plating whenever possible.

Crucially, even in systems where Li losses are self-limiting, a substantial fraction of Li becomes irreversibly trapped within the first few cycles. This rapid depletion of active ion presents a key challenge for anode-free batteries (also referred to as anode-less or zero-excess designs), which rely on Li plating to serve as the functional anode. As a result, such architectures remain highly challenging for conventional liquid-electrolyte cells and continue to face significant limitations even in solid-state systems, given their susceptibility to irreversible Li trapping and cumulative loss mechanisms. Nevertheless, advances in electrolyte and interface engineering[23] demonstrate that partial mitigation is possible, highlighting the need to address irreversible Li trapping as a central limitation for the practical implementation of anode-free batteries.

**Experimental Methods**

The experimental methodology closely follows that described in our previous work,[11] with minor modifications. All metal components were used as 25 µm-thick foils with a purity above 98% (supplied by Goodfellow). The components were cut into 12-mm diameter circular electrodes using a high-precision cutter (EL-Cut, EL-CELL). Electrochemical cells were assembled in a Swagelok-type two-terminal configuration. Each metal electrode was separated from the Li source (99.9%-pure metal Li foil, S4R) using a Whatman GF/B glass microfiber filter in combination with a Whatman qualitative filter paper, the latter preventing fiber contamination of the metal surface. Each separator was moistened with 20 µL of 1.0 M $LiPF_6$ electrolyte in a solvent mixture of ethylene carbonate, diethyl carbonate and dimethyl carbonate (EC/DEC/DMC with a volume ratio 1:1:1, by Sigma Aldrich).

Their electrochemical performances were investigated after a period of 8 hours from assembly, monitoring the $V_{oc}$ to ensure its stabilization. To investigate different lithiation kinetic regimes, galvanostatic charge-discharge (GCD) experiments were performed at constant currents within the range of 20 µA—2 mA, at a C/2 rate. The voltage window used in these cycles was −1.0 to 3.0 V. Therefore, each cycle was terminated when either the target capacity or the voltage limits were reached. Electrochemical tests were conducted using Gamry Interface 1010e potentiostats. After lithiation, metal components were extracted from the cells and rinsed with battery-grade DMC (Sigma-Aldrich) for post-mortem studies. The sample preparation, extraction and mounting in the corresponding sample holders (adapted for each experiment) were performed in an Ar-filled Jacomex glovebox. The samples were transported to the analysis setups in sealed aluminum bags.

Ion beam analysis (IBA), including nuclear reaction analysis (NRA) and Rutherford backscattering spectrometry (RBS), was conducted at the Centro de Microanálisis de Materiales de Madrid (CMAM, UAM, Madrid, Spain)[24] using the same experimental setup as previously reported. NRA was performed with currents of 16—18 nA and RBS with 24—30 nA, each until cumulative charges of 10 µC were reached. The resulting NRA and RBS spectra were analyzed using the MultiSIMNRA[17] computational tool (version 1.6), based on SIMNRA 7.04,[25,26] with collisional cross-sections as reported by Paneta et al.[18]

**Supporting Information Available**

Galvanostatic Charge-Discharge (GCD) testing including intermediate currents, Ion Beam Analysis (IBA) data of cycled metal components, lithiation/delithiation kinetics in Ag electrodes, residual Li capacity described through power-law function, and residual capacity and irreversibility analysis in Mg electrodes (PDF).

**Acknowledgements**

This work was supported by the European Union through the Horizon Europe Research and Innovation Program under grant agreement No. 101103834 (Project OPERA). It also received funding from the Spanish national Agencia Estatal de Investigación (AEI) as part of the following programs: M-ERA.NET Call 2021 (Project SOLIMEC, subprojects PCI2022-132955 and PCI2022-132998), "Proyectos de Generación de Conocimiento" (Project NanoCatCom, Ref. PID2021–1246670B) and "María de Maeztu" Programme for Units of Excellence in R&D (CEX2023-001316-M). We acknowledge the service from the MiNa Laboratory at IMN, and funding from CM (project S2018/NMT-4291 TEC2SPACE), MINECO (project CSIC13-4E-1794) and EU (FEDER, FSE). We also acknowledge the support from the Center for Micro-Analysis of Materials (CMAM) - Universidad Autónoma de Madrid, for the beam time proposals, with codes STD011/24 and STD032/24, and its technical staff for their contribution to the operation of the accelerator.

**Author Information**

| | | |
|---|---|---|
| Arturo Galindo | (arturo.galindo.sanz@csic.es) | ORCID: 0000-0003-2030-2818 |
| Jesús Diaz-Sánchez | (jesus.diazs@uam.es) | ORCID: 0000-0002-3780-3726 |
| Sunil Kumar | (sunil.kumar@uam.es) | ORCID: 0000-0002-6542-2760 |
| Bouthayna Alrifai | (bouthayna.alrifai@uam.es) | ORCID: 0000-0002-5353-6592 |
| Andrea Marchetti | (andrea.marchetti@csic.es) | ORCID: 0009-0008-6805-0000 |
| Gastón García | (gaston.garcia@uam.es) | ORCID: 0000-0002-1866-4871 |
| Celia Polop | (celia.polop@uam.es) | ORCID: 0000-0003-3607-7643 |
| Enrique Vasco | (enrique.vasco@csic.es) | ORCID: 0000-0002-0647-4499 |

The authors state that there are no conflicts of financial/commercial interest or personal relationships that could have influenced the work reported in this article.

**Author contributions**

**A. Galindo:** Investigation, Methodology, Formal analysis, Visualization. **J. Diaz-Sanchez:** Investigation, Methodology. **S. Kumar:** Resources, Validation. **B. Alrifai:** Methodology, Writing– review &editing. **A. Marchetti:** Methodology. **G. García:** Formal analysis, Writing– review &editing. **C. Polop:** Writing– review & editing, Project administration, Funding acquisition. **E. Vasco:** Writing– original draft, Conceptualization, Supervision, Project administration, Funding acquisition.

**References**

(1) Tarascon, J.-M.; Armand, M. Issues and Challenges Facing Rechargeable Lithium Batteries. *Nature* **2001**, *414*, 359-367. https://doi.org/10.1038/35104644.

(2) Choi, J. W.; Aurbach, D. Promise and Reality of Post-Lithium-Ion Batteries with High Energy Densities. *Nat. Rev. Mater.* **2016**, *1* (4), 16013. https://doi.org/10.1038/natrevmats.2016.13.

(3) Vetter, J.; Novák, P.; Wagner, M. R.; Veit, C.; Möller, K. C.; Besenhard, J. O.; Winter, M.; Wohlfahrt-Mehrens, M.; Vogler, C.; Hammouche, A. Ageing Mechanisms in Lithium-Ion Batteries. *J. Power Sources* **2005**, *147* (1–2), 269–281. https://doi.org/10.1016/j.jpowsour.2005.01.006.

(4) Andre, D.; Kim, S.-J.; Lamp, P.; Lux, S. F.; Maglia, F.; Paschos, O.; Stiaszny, B. Future Generations of Cathode Materials: An Automotive Industry Perspective. *J. Mater. Chem. A* **2015**, *3* (13), 6709–6732. https://doi.org/10.1039/C5TA00361J.

(5) Xie, W.; Liu, X.; He, R.; Li, Y.; Gao, X.; Li, X.; Peng, Z.; Feng, S.; Feng, X.; Yang, S. Challenges and Opportunities toward Fast-Charging of Lithium-Ion Batteries. *J. Energy Storage* **2020**, *32*, 101837. https://doi.org/10.1016/j.est.2020.101837.

(6) Keil, P.; Schuster, S. F.; Wilhelm, J.; Travi, J.; Hauser, A.; Karl, R. C.; Jossen, A. Calendar Aging of Lithium-Ion Batteries. *J. Electrochem. Soc.* **2016**, *163* (9), A1872–A1880. https://doi.org/10.1149/2.0411609jes.

(7) Wei, Z.; Yang, X.; Li, Y.; He, H.; Li, W.; Sauer, D. U. Machine Learning-Based Fast Charging of Lithium-Ion Battery by Perceiving and Regulating Internal Microscopic States. *Energy Storage Mater.* **2023**, *56*, 62–75. https://doi.org/10.1016/j.ensm.2022.12.034.

(8) Ji, G.; He, L.; Wu, T.; Cui, G. The Design of Fast Charging Strategy for Lithium-Ion Batteries and Intelligent Application: A Comprehensive Review. *Appl. Energy* **2025**, *377* (Part C), 124538. https://doi.org/10.1016/j.apenergy.2024.124538.

(9) Burke, A.; Miller, M. The Power Capability of Ultracapacitors and Lithium Batteries for Electric and Hybrid Vehicle Applications. *J. Power Sources* **2011**, *196* (1), 514–522. https://doi.org/10.1016/j.jpowsour.2010.06.092.

(10) Offer, G. J.; Howey, D.; Contestabile, M.; Clague, R.; Brandon, N. P. Comparative Analysis of Battery Electric, Hydrogen Fuel Cell and Hybrid Vehicles in a Future Sustainable Road Transport System. *Energy Policy* **2010**, *38* (1), 24–29. https://doi.org/10.1016/j.enpol.2009.08.040.

(11) Galindo, A.; Xavier, N.; Maldonado, N.; Díaz-Sánchez, J.; Morant, C.; García, G.; Polop, C.; Cai, Q.; Vasco, E. Lithiation Analysis of Metal Components for Li-Ion Battery Using Ion Beams. *Battery Energy* **2025**, *5* (1), e70076. https://doi.org/10.1002/bte2.70076.

(12) Okamoto, H. . *Desk Handbook: Phase Diagrams for Binary Alloys*. **2010**, 855.

(13) Okamoto, H. Al-Li (Aluminum-Lithium). *J. Phase Equilibria Diffus.* **2012**, *33* (6), 500–501. https://doi.org/10.1007/s11669-012-0119-8.

(14) Okamoto, H. Supplemental Literature Review of Binary Phase Diagrams: Cs-In, Cs-K, Cs-Rb, Eu-In, Ho-Mn, K-Rb, Li-Mg, Mg-Nd, Mg-Zn, Mn-Sm, O-Sb, and Si-Sr. *J. Phase Equilibria Diffus.* **2013**, *34* (3), 251–263. https://doi.org/10.1007/s11669-013-0233-2.
20